\documentclass{ifacconf}

\usepackage{graphicx}      
\usepackage{natbib}        

\usepackage{amsmath}
\usepackage{amssymb}
\usepackage{amsfonts}
\usepackage{graphicx}
\usepackage{lipsum}
\usepackage{epstopdf}
\usepackage{algorithmic}
\usepackage{mathptmx}
\ifpdf
  \DeclareGraphicsExtensions{.eps,.pdf,.png,.jpg}
\else
  \DeclareGraphicsExtensions{.eps}
\fi
\DeclareMathAlphabet{\bit}{OML}{cmm}{b}{it}

\def\fB{\mathfrak{B}}

\def\<{\leqslant}           
\def\>{\geqslant}           

\def\d{\partial}
\def\wh{\widehat}

\def\cT{\mathcal{T}}   

\def\mR{\mathbb{R}}    
\def\mC{\mathbb{C}}    

\def\Tr{\mathrm{Tr}}       
\def\rT{\mathrm{T}}        
\def\cov{\mathbf{cov}}
\def\var{\mathbf{var}}

\def\bE{\mathbf{E}}    

\def\bP{\mathbf{P}}    

\def\[[[{[\![\![}   
\def\]]]{]\!]\!]}   

\def\bra{{\langle}}
\def\ket{{\rangle}}

\def\re{\mathrm{e}}        
\def\rd{\mathrm{d}}        

\def\x{\times}

\def\fF{\mathfrak{F}}

\def\fS{\mathfrak{S}}

\def\cW{\mathcal{W}}

\def\cI{\mathcal{I}}

\def\mS{\mathbb{S}}

\def\Ups{\Upsilon}

\def\endpf{\hfill$\blacksquare$}

\begin{document}
\pagestyle{plain}
\begin{frontmatter}

\title{
Filtering and $1/3$ Power Law for Optimal Time Discretisation
in
Numerical Integration of Stochastic Differential Equations\thanksref{footnoteinfo}}

\thanks[footnoteinfo]{This work is supported by the Australian Research Council grant DP240101494.}

\author[I]{Igor G. Vladimirov}

\address[I]{School of Engineering, Australian National University, Canberra, ACT 2601, Australia (e-mail: igor.g.vladimirov@gmail.com)}

\begin{abstract}                
This paper is concerned with the numerical integration of stochastic differential equations (SDEs) which govern diffusion processes driven by a standard Wiener process. With the latter being replaced by a sequence of increments at discrete moments of time, we revisit a filtering point of view on the approximate strong solution of the SDE as an estimate of the hidden system state whose conditional  probability distribution is updated using a Bayesian approach and Brownian bridges over the intermediate time intervals. For a class of multivariable linear SDEs, where the numerical solution is organised as a Kalman filter, we investigate the fine-grid asymptotic behaviour of terminal and integral mean-square error functionals when the time discretisation is specified by a sufficiently smooth monotonic transformation of a uniform grid. This leads to constrained optimisation problems over the time discretisation profile, and their solutions reveal a $1/3$ power law for the asymptotically optimal grid density functions. As a one-dimensional  example, the results are illustrated for the Ornstein-Uhlenbeck process.
\end{abstract}

\begin{keyword}
stochastic differential equation,
numerical integration,
time discretisation,
stochastic filtering,
Brownian bridge,
Kalman filter,
mean-square error functional,
grid optimisation.
\end{keyword}

\end{frontmatter}

\section{Introduction}

The dynamics of continuous-time physical systems subject to random environments are often modelled by stochastic differential equations (SDEs) of Ito type,  which govern a diffusion process of interest and are driven by a standard Wiener process representing the external noise acting on the system. Numerical integration of SDEs is carried out in Monte Carlo simulations of such systems and is particularly important in situations where of relevance is not only an accurate reproduction of the probability law of the process but also its pathwise dependence on the driving noise.
A number of computational schemes, pioneered by Mil'shtein's method [\cite{M_1974,M_1978}],   were developed over decades for the numerical solution of SDEs and are employed both in natural sciences and for modelling artificial systems such as financial markets [\cite{KP_1992}]. In the numerical integration of an SDE, the driving Wiener process is usually replaced with a sequence of its increments at a sufficiently dense grid of points in  the simulation time interval,  and the SDE itself is approximated by a difference equation.  The quality of the resulting discrete-time approximation of the underlying diffusion process is quantified in terms of mean-square error functionals (or other deviation measures) and their asymptotic behaviour, including the order  of convergence to zero   together with the grid stepsize.
The approximate strong solution of the SDE can be looked at from a stochastic filtering standpoint [\cite{N_1991}] as an estimate  of the hidden system state based on its initial condition and the available Wiener process increments as an observation history. The corresponding posterior probability distribution of the diffusion process is then updated using a Bayesian approach and Brownian bridges resulting from conditioning the driving Wiener process on its discrete-time increments over the intermediate time intervals.
After a brief review of the stochastic numerics and Bayesian filtering in Sections~\ref{numer} and \ref{bridge}, the present paper proceeds in Section~\ref{linear} to a Kalman filter implementation of the mean-square optimal numerical solution for multivariable linear SDEs, which are principal models  in  linear-quadratic Gaussian (LQG) filtering and control theory [\cite{AM_1989,KS_1972}]. Then in Sections~\ref{prof} and \ref{MSfun}, we consider a class of time discretisations,  obtained by a continuously differentiable monotonic transformation of a uniform grid,  and investigate the fine-grid asymptotic behaviour of terminal and integral mean-square error functionals for the corresponding numerical solution of the linear SDE. The minimisation of the leading coefficients in these asymptotic relations over the time discretisation profile gives rise to constrained optimization problems in Section~\ref{opt},  where the grid density (as a function of time) plays the role of a control strategy.  Their solution yields asymptotically optimal grid densities whose dependence on the controllability and observability Gramians  reveals a $1/3$ power law. We illustrate these results (including a comparison with the case of uniform grids) for an  Ornstein-Uhlenbeck process in Section~\ref{example} and provide concluding remarks in Section~\ref{conc}.

\section{Numerics of Time invariant stochastic systems}
\label{numer}

Consider a stochastic system (for simplicity, time invariant) whose state is an $\mR^n$-valued diffusion process $X:= (X_t)_{t\> 0}$ evolving in time $t$ according to an Ito SDE
\begin{equation}
\label{SDE}
  \rd X_t = f(X_t)\rd t + g(X_t) \rd W_t,
  \qquad
  t\> 0.
\end{equation}
It is driven by a standard Wiener process $W:= (W_t)_{t\> 0}$ in $\mR^m$ adapted to a filtration $(\fF_t)_{t\> 0}$ of an underlying probability space $(\Omega, \fF, \bP)$ satisfying the usual assumptions  [\cite{KS_1991}], with the initial condition  $X_0$ being $\fF_0$-measurable (so that $X_0$ and $W$ are independent).
The maps $f: \mR^n \to \mR^n$ and $g:  \mR^n \to \mR^{n\x m}$ describe the drift vector and the dispersion matrix, respectively. They are assumed to be sufficiently smooth and uniformly Lipschitz continuous in $\mR^n$  (and thus with a sublinear growth at infinity)
in order to guarantee the existence and uniqueness of strong solutions of (\ref{SDE}) without a finite-time blow-up.
Similarly to ODEs, computational methods for numerical solution of the SDE (\ref{SDE}) are derived from its equivalent integral representation
\begin{equation}
\label{SDEint}
    X_t
    =
    X_s + \int_s^t f(X_u) \rd u + \int_s^t g(X_u) \rd W_u,
    \qquad
    t \> s \> 0,
\end{equation}
on an interval $[s,t]$. A typical numerical method is obtained by successively  applying the representation (\ref{SDEint}) to $N$ time intervals
\begin{equation}
\label{Deltak}
    \Delta_k := [t_k, t_{k+1}],
    \qquad
    0 \< k < N,
\end{equation}
of lengths $\Delta t_k:= t_{k+1}-t_k>0$,
associated with the discretisation
\begin{equation}
\label{ttt}
    0 = t_0 < t_1 < \ldots < t_N = T
\end{equation}
of the interval $[0,T]$ for a given time horizon $T>0$,  and its accuracy is studied asymptotically, as $N\to +\infty$ together with $\max_{0\< k < N} \Delta t_k \to 0$.    For example, the Euler-Maruyama method [\cite{KP_1992}] approximates the corresponding values $(X_{t_k})_{0\< k \< N}$ of the process $X$ (which form a Markov chain) using a difference equation
\begin{align}
\nonumber
    X_{t_{k+1}}
    & :=
      X_{t_k}
      +
    \int_{\Delta_k} f(X_t) \rd t + \int_{\Delta_k} g(X_t) \rd W_t    \\
\label{EM}
    & \approx
    X_{t_k}
    +
    f(X_{t_k})
    \Delta t_k
    +
    g(X_{t_k})
    \Delta W_k,
\end{align}
based on considering the drift vector $f(X_t)$ and the dispersion matrix $g(X_t)$ to be constant over the time interval $\Delta_k$. The increments
\begin{equation}
\label{DeltaWk}
    \Delta W_k
    :=
    W_{t_{k+1}} - W_{t_k}
\end{equation}
of the standard Wiener process $W$
are zero-mean Gaussian random vectors in $\mR^m$, which are mutually independent (and also independent of the initial condition $X_0$) and have the covariance matrices
\begin{equation}
\label{covDeltaWk}
    \cov(\Delta W_k)
    =
    I_m \Delta t_k,
\end{equation}
where $I_m$ is the identity matrix of order $m$.
Compared to (\ref{EM}),  an asymptotically  more accurate alternative is provided by Mil'shtein's method [\cite{M_1974,M_1978}]:
\begin{align}
\nonumber
    X_{t_{k+1}}
    \approx &
    X_{t_k}
    +
    f(X_{t_k})
    \Delta t_k
    +
    g(X_{t_k})
    \Delta W_k\\
\label{MM}
    & +
    \sum_{\ell = 1}^n
    h_\ell(X_{t_k})
    \int_{\Delta_k}
    (\rd W_t)
    (W_t-W_{t_k})^\rT
    (g_{\ell \bullet}(X_{t_k}))^\rT,
\end{align}
where the maps
\begin{equation}
\label{hell}
    h_\ell := \d_{x_\ell}g : \mR^n \to \mR^{n\x m},
    \qquad
    \ell = 1, \ldots, n,
\end{equation}
are the partial derivatives of the dispersion matrix map $\mR^n \ni x:= (x_\ell)_{1\< \ell \< n}  \mapsto g(x) \in \mR^{n\x m}$ with respect to the entries of the vector $x$, and $g_{\ell \bullet}: \mR^n\to \mR^{1\x m}$ denotes the $\ell$th row of the matrix:
\begin{equation}
\label{gg}
    g = \begin{bmatrix}
      g_{1\bullet}\\
      \vdots \\
      g_{n \bullet}
    \end{bmatrix}.
\end{equation}
The rightmost sum in (\ref{MM})  can also be represented as a vector
$    \sum_{\ell = 1}^n
    h_\ell(X_{t_k})
    \int_{\Delta_k}
    (\rd W_t)
    (W_t-W_{t_k})^\rT
    (g_{\ell \bullet}(X_{t_k}))^\rT
     =
    (
    \int_{\Delta_k}
    (W_t-W_{t_k})^\rT H_j(X_{t_k}) \rd W_t
    )_{1\< j \< n}
$ of bilinear  forms of $W$,
where the matrix-valued maps $H_j: \mR^n \to \mR^{m\x m}$ are expressed in terms of (\ref{hell}), (\ref{gg}) as
$    H_j = \sum_{\ell =1}^n
    g_{\ell \bullet}^\rT (h_\ell)_{j \bullet}$ for all $j = 1, \ldots, n$.
If the latter take values in the subspace $\mS_m \subset \mR^{m\x m}$ of real symmetric matrices of order $m$, then
$
        \int_{\Delta_k}
    (W_t-W_{t_k})^\rT H_j\rd W_t
    =
    \frac{1}{2}
    (\Delta W_k^\rT H_j \Delta W_k - \Tr H_j \Delta t_k  )
$,
where the argument $X_{t_k}$ is omitted for brevity, and use is made of the stochastic differential
$
    \rd (W_t^\rT  H W_t) = 2 W_t^\rT H \rd W_t + \Tr H \rd t
$
(with a constant matrix $H \in \mS_m$) obtained through the Ito lemma [\cite{KS_1991}]. The Mil'shtein scheme admits further refinements which employ multiple Wiener integrals [\cite{KP_1992}] arising from higher-order Taylor approximations of the integrands $f$ and $g$ and iterations of the integral operators on the right-hand sides of (\ref{SDEint}), (\ref{EM}). The accuracy of such methods at any time step $k$ is quantified by the discrepancy (for example, a mean-square error) between $X_{t_{k+1}}$ and its numerical approximation $\wh{X}_{t_{k+1}}$, produced from the initial condition $X_0$ and the increments $\Delta W_0, \ldots,  \Delta W_k$ in (\ref{DeltaWk}). The latter represent the continuous time driving process $W$ only at the discrete moments of time $t_1, \ldots, t_{k+1}$ (with $W_0 = 0$), while its intermediate behaviour on the intervals (\ref{Deltak}) remains unobservable.

\section{Intermediate Brownian bridges and filtering}
\label{bridge}

The conditional distribution of the standard Wiener process $W$ on the time interval $\Delta_k$ in (\ref{Deltak}),  given its increments $\Delta W_0, \ldots, \Delta W_k$ in (\ref{DeltaWk}),  is Gaussian, and, due to the Markov property of $W$,  depends only on $W_{t_k} = \sum_{j=0}^{k-1} \Delta W_j$ and $\Delta W_k$ through the conditional mean and covariance function:
\begin{align}
\label{EWt}
    \bE(W_t  \mid  \cW_k)
    & =
    W_{t_k}
    + \frac{t-t_k}{\Delta t_k}\Delta W_k,\\
\label{covWst}
    \cov(W_s, W_t   \mid  \cW_k)
    & =
    \Big(\!\!
    \min(s,t)-t_k
    -
    \frac{(s-t_k)(t-t_k)}{\Delta t_k}
    \Big)
    I_m
\end{align}
for all $s,t \in \Delta_k$. Here, use is made of a Gaussian random vector
\begin{equation}
\label{cWk}
  \cW_k := (\Delta W_j)_{0\< j \< k}
\end{equation}
and the normal correlation lemma [\cite{LS_2001}] along with the relations $\bE W_u = 0$ and $\cov(W_u, W_v) = \min(u,v)I_m$ for any $u, v \> 0$.
The conditionally Gaussian law, specified by (\ref{EWt}), (\ref{covWst}),  corresponds to an $\mR^m$-valued    Brownian bridge [\cite{RW_2000}] with an increment $\Delta W_k$ over time $\Delta t_k$. This leads to a recurrence equation for the conditional probability distribution
\begin{equation}
\label{Pk}
      P_k(B)
    :=
    \bP(X_{t_{k+1}} \in B \mid \fS_k),
    \qquad
    B \in \fB^n,
\end{equation}
of the  random vector $X_{t_{k+1}}$ on the $\sigma$-algebra $\fB^n$ of Borel subsets of $\mR^n$ with respect to the $\sigma$-algebra
\begin{equation}
\label{fSk}
  \fS_k
  :=
  \sigma\{X_0, \cW_k\}
\end{equation}
of events
generated by the initial system state $X_0$ and the vector (\ref{cWk}), with $\fS_{-1}:= \sigma\{X_0\}$.  Note that $\fS_k \subset \fF_{t_{k+1}}$, whereby $\Delta W_k$ in (\ref{DeltaWk}) is independent of $\fS_{k-1}$.
The conditional expectation
\begin{equation}
\label{EXtk}
    \wh{X}_{t_{k+1}}
    :=
    \bE(X_{t_{k+1}}\mid \fS_k)
    =
    \int_{\mR^n}
    x P_k(\rd x)
\end{equation}
provides a mean-square optimal estimator of $X_{t_{k+1}}$ in the class of $\mR^n$-valued  Borel functions of $X_0$ and $\cW_k$, with the latter playing  the role of an observation history. In order to formulate the recurrence below,
for any $\tau> 0$, $x \in \mR^n$ and $w \in \mR^m$,  let $\Pi_{\tau, x, w}$ denote the probability distribution which $X_\tau$ would have if the SDE (\ref{SDE}) were initialised with $X_0= x$ and driven  on the time interval $[0,\tau]$ by an $\mR^m$-valued  Brownian bridge $\zeta := (\zeta_t)_{0\< t \< \tau}$  (instead of the standard Wiener process) with the increment $w$, so that
\begin{equation}
\label{zeta0tau}
    \zeta_0 = 0,
    \qquad
    \zeta_\tau = w.
\end{equation}
More precisely, $\zeta$ is a diffusion process satisfying a time-varying SDE
\begin{equation}
\label{SDEbridge}
    \rd \zeta_t = \frac{1}{t-\tau} (\zeta_t - w)\rd t + \rd \omega_t,
    \qquad
    0 \< t \< \tau,
\end{equation}
driven by an  $\mR^m$-valued standard Wiener process $\omega := (\omega_t)_{t \> 0}$.  Accordingly, $\Pi_{\tau, x, w}$ is the distribution of  the terminal value $\xi_\tau$ of an auxiliary Ito process $\xi:= (\xi_t)_{0 \< t \< \tau}$ in $\mR^n$ governed by the SDE
\begin{equation}
\label{SDExi}
    \rd \xi_t = f(\xi_t)\rd t + g(\xi_t) \rd \zeta_t,
    \qquad
    0 \< t \< \tau,
\end{equation}
with the initial condition $\xi_0 = x$. A combined form of (\ref{SDEbridge}), (\ref{SDExi}) is
$$
    \rd
    \begin{bmatrix}
        \xi_t\\
        \zeta_t
    \end{bmatrix}
    =
    \begin{bmatrix}
        f(\xi_t) + \frac{1}{t-\tau} g(\xi_t) (\zeta_t-w)\\
        \frac{1}{t-\tau} (\zeta_t-w)
    \end{bmatrix}
    \rd t
    +
    \begin{bmatrix}
      g(\xi_t)\\
      I_m
    \end{bmatrix}
    \rd \omega_t,
$$
which governs a diffusion process in the augmented space $\mR^{n+m}$ initialised at ${\small\begin{bmatrix}
  x\\
  0
\end{bmatrix}}$  and landing at time $\tau$ on an affine subspace specified by $w$ in (\ref{zeta0tau}). The probability  measure $\fB^n \ni B \mapsto \Pi_{\tau, x, w}(B) = \bP(\xi_\tau \in B)$ provides a transition kernel for
the conditional distributions (\ref{Pk}):
\begin{equation}
\label{PkB}
    P_k(B)
    =
    \int_{\mR^n}
    \Pi_{\Delta t_k, x, \Delta W_k}(B)
    P_{k-1}(\rd x),
    \qquad
    0 \< k < N.
\end{equation}
This relation is a recursive filtering equation for updating the posterior distribution $P_k$ of the hidden system state $X_{t_{k+1}}$ based on the initial condition $X_0$ and the observation history $\cW_k$ from (\ref{cWk}). As a filter, the resulting numerical solver of the SDE (\ref{SDE}) produces all the information which the discrete time data carry about $X_{t_{k+1}}$, including the conditional means (\ref{EXtk}).
 Its performance depends on the grid (\ref{ttt}),  and the choice of the latter is an  observation control problem which can be endowed with optimality criteria.

\section{Numerical integration of linear SDEs via Kalman filter}
\label{linear}

As mentioned in Introduction, LQG filtering and control theory
uses
the  class of stochastic systems (\ref{SDE}) with linear   drift maps $  f(x) := Ax$ and constant dispersion matrices $g(x):= B$ for all $x \in \mR^n$,
with $A \in \mR^{n\x n}$ and $B \in \mR^{n\x m}$,  leading to a linear SDE:
\begin{equation}
\label{SDElin}
    \rd X_t = AX_t + B\rd W_t.
\end{equation}
Due to this  linearity, the exact part of (\ref{EM}) lends itself to a closed-form representation:
\begin{equation}
\label{dXklin}
    X_{t_{k+1}}
    =
    \re^{A \Delta t_k} X_{t_k}
    +
    Z_k,
    \qquad
    0 \< k < N.
\end{equation}
Here,
\begin{equation}
\label{Zk}
  Z_k
  :=
      \int_{\Delta_k}
    \re^{(t_{k+1}-t) A}
    B \rd W_t
\end{equation}
are $\mR^n$-valued zero-mean Gaussian random vectors, which are mutually independent  together with $X_0$ (with $Z_k$ being independent of $X_{t_k}$ as well) and have the following covariance matrices:
\begin{equation}
\label{covZk}
  \cov(Z_k)
  =
      \int_{\Delta_k}
    \re^{(t_{k+1}-t) A}
    D
    \re^{(t_{k+1}-t) A^\rT}
    \rd t
    =
    G_{\Delta t_k}.
\end{equation}
These matrices are expressed in terms of the  diffusion matrix
\begin{equation}
\label{D}
  D := BB^\rT
\end{equation}
of the SDE (\ref{SDElin}) and the finite-horizon controllability Gramian of the pair $(A,B)$:
\begin{align}
\nonumber
  G_t
  & :=
  \int_0^t \re^{sA} D \re^{sA^\rT} \rd s\\
\label{G}
  & =
  t
  \sum_{j,k\> 0}
  \frac{1}{j!k!(j+k+1)}
  A^j D (A^\rT)^k
  t^{j+k} ,
  \qquad
  t \> 0.
\end{align}
The latter satisfies the initial value problem for the Lyapunov ODE
\begin{equation}
\label{Gdot}
    \dot{G}_t = A G_t + G_t A^\rT + D,
    \qquad
    G_0 = 0,
\end{equation}
where $\dot{(\ )}$ is the time derivative.
In the linear case being considered, the conditional distributions $P_k$ in (\ref{Pk}) are Gaussian,  and the recurrence equation (\ref{PkB}) takes the form of a Kalman filter [\cite{LS_2001}] which updates the conditional means (\ref{EXtk}) and covariance matrices according to the relations
\begin{align}
\nonumber
    \mu_k
    := &
    \wh{X}_{t_{k+1}}
    =
    \bE (X_{t_{k+1}} \mid \fS_{k-1})
    +
    \cov (X_{t_{k+1}}, \Delta W_k \mid \fS_{k-1})\\
\nonumber
    & \x
    \cov (\Delta W_k \mid \fS_{k-1})^{-1} (\Delta W_k - \bE (\Delta W_k \mid \fS_{k-1}))\\
\nonumber
    = &
    \re^{A \Delta t_k}
    \mu_{k-1}
    +
    \cov(Z_k,\Delta W_k)
    \cov(\Delta W_k)^{-1}
    \Delta W_k\\
\label{muk}
    = &
    \re^{A \Delta t_k}
    \mu_{k-1}
    +
    E(A \Delta t_k)B \Delta W_k, \\
\nonumber
    \Sigma_k
    := &
    \cov(X_{t_{k+1}}\mid \fS_k)\\
\nonumber
    = &
    \cov(X_{t_{k+1}}\mid \fS_{k-1})
    -
    \cov (X_{t_{k+1}}, \Delta W_k \mid \fS_{k-1})\\
\nonumber
    & \x
    \cov (\Delta W_k \mid \fS_{k-1})^{-1}
    \cov (X_{t_{k+1}}, \Delta W_k   \mid \fS_{k-1})^\rT\\
\nonumber
    = &
    \re^{A \Delta t_k}
    \Sigma_{k-1}
    \re^{A^\rT \Delta t_k}
    +
    \cov(Z_k)\\
\nonumber
    & -
    \cov(Z_k,\Delta W_k)
    \cov(\Delta W_k)^{-1}
    \cov(Z_k,\Delta W_k)^\rT\\
\nonumber
    = &
    \re^{A \Delta t_k}
    \Sigma_{k-1}
    \re^{A^\rT \Delta t_k}
    +
    G_{\Delta t_k}-
    E(A \Delta t_k) D E(A^\rT \Delta t_k)\Delta t_k\\
\label{Sigmak}
    = &
    \re^{A \Delta t_k}
    \Sigma_{k-1}
    \re^{A^\rT \Delta t_k}
    +K_{\Delta t_k}(\Delta t_k)^3,
    \qquad
    0 \< k < N,
\end{align}
which are initialised at
\begin{equation}
\label{muSigma0}
    \mu_{-1}
    :=
    \bE(X_0 \mid X_0)
    =
    X_0,
    \qquad
    \Sigma_{-1}
    :=
    \cov(X_0 \mid X_0 )
    =
    0
\end{equation}
in accordance with (\ref{fSk}). Here, use is made of (\ref{covDeltaWk}), (\ref{dXklin}). Also,
\begin{equation}
\label{E}
  E(z)
  :=
  \int_0^1
  \re^{sz}
  \rd s
  =
  \frac{\re^z-1}{z}
  =
  \sum_{k \> 0}
  \frac{z^k}{(k+1)!},
  \qquad
  z \in \mC,
\end{equation}
 is an entire function
of a complex variable (with  $E(0) = 1$). It is
evaluated [\cite{H_2008}] in (\ref{muk}) at the matrix $A \Delta t_k$ and arises
from
$
    \cov(Z_k, \Delta W_k)
     =
    \bE
    (
    \int_{\Delta_k}
    \re^{(t_{k+1}-s) A}
    B
    \rd W_s
    \int_{\Delta_k}
    \rd W_t^\rT
    )
     =
    \int_0^{\Delta t_k}
    \re^{tA}
    B
    \rd t
    =
    E(A \Delta t_k) B \Delta t_k
$,
which is obtained from (\ref{Deltak}), (\ref{Zk}) and, together with (\ref{covZk})--(\ref{G}),  is also used in (\ref{Sigmak}). The rightmost term in (\ref{Sigmak}) involves a function
\begin{align}
\nonumber
    & K_t
    :=
    \frac{1}{t^2}
    \Big(
    \frac{1}{t}
    G_t-
    E(tA) D E(tA^\rT)
    \Big)\\
\nonumber
     & =
    \frac{1}{t^2}
    \sum_{j,k\> 0}
    \Big(
  \frac{1}{j!k!(j+k+1)}
  -
  \frac{1}{(j+1)!(k+1)!}
  \Big)
  A^j D (A^\rT)^k
  t^{j+k}\\
\label{K}
    & =
    \sum_{j,k\> 1}
  \frac{jk\, t^{j+k-2}}{(j+1)!(k+1)!(j+k+1)}
  A^j D (A^\rT)^k,
  \qquad
  t> 0,\!\!
\end{align}
which takes values in the set $\mS_n^+ \subset \mS_n$ of real positive semi-definite symmetric matrices of order $n$, including the limit value
\begin{equation}
\label{mho}
  \mho
  :=
  \lim_{t\to 0+}
  K_t
  = \frac{1}{12} ADA^\rT.
\end{equation}
The series in (\ref{K}) inherits from (\ref{G}), (\ref{E}) the infinite radius of convergence with respect to $t$.  Since the conditional expectations (\ref{muk}) constitute a discrete time  numerical solution of the SDE (\ref{SDElin}), we will be concerned with the conditional covariance matrices (\ref{Sigmak}) which quantify the mean-square accuracy of this solution at all steps. In particular, for a given matrix $M \in \mS_n^+$, which specifies the relative importance of the system variables,
\begin{align}
\nonumber
    \bE (\|X_{t_{k+1}}-\mu_k\|_M^2)
    & =
    \bE
    \bE (\|X_{t_{k+1}}-\mu_k\|_M^2 \mid \fS_k) \\
\label{MSigma}
    & =
    \bra M, \bE \Sigma_k\ket
    =
    \bra M, \Sigma_k\ket,
\end{align}
where $\|v\|_M:= \sqrt{v^\rT M v} = |\sqrt{M}v|$ is a weighted Euclidean norm of a vector $v \in \mR^n$, and $\bra K,L\ket:= \Tr (K^\rT L)$ is the Frobenius inner product [\cite{HJ_2007}] of equally dimensioned real matrices $K$, $L$. In (\ref{MSigma}), use is also made of the tower property of conditional expectations along with the nonrandomness of conditional covariance matrices in the Gaussian case [\cite{LS_2001}]. In accordance with (\ref{ttt}), the \emph{terminal} mean-square error of (\ref{muk}) as a numerical solution of (\ref{SDElin})   is given by (\ref{MSigma}) with $k=N-1$:
\begin{equation}
\label{MSfin}
    \cT_N
    :=
    \bE (\|X_T-\mu_{N-1}\|_M^2)
    =
    \bra M, \Sigma_{N-1}\ket.
\end{equation}
An alternative mean-square functional of \emph{integral} type is provided by
\begin{equation}
\label{MSint}
    \cI_N
    :=
    \bE
    \sum_{k=0}^{N-1}
    \|X_{t_{k+1}}-\mu_k\|_M^2\Delta t_k
    =
        \sum_{k=0}^{N-1}
    \bra
        M,
        \Sigma_k
    \ket
    \Delta t_k
\end{equation}
and quantifies the performance of (\ref{muk}) over the interval $[0,T]$. For example, if $M = I_n$, then (\ref{MSigma})--(\ref{MSint}) involve the conditional variances    $\bra I_n, \Sigma_k\ket = \Tr \Sigma_k = \var(X_{t_{k+1}} \mid \fS_k)$.

\section{Time discretisation profile and asymptotics of covariance matrices}
\label{prof}

For a fixed but otherwise arbitrary horizon $T>0$, suppose the discretisation of the time interval $[0,T]$ in (\ref{ttt}) is obtained by a monotonic transformation of the uniform grid of the unit interval:
\begin{equation}
\label{tttphi}
  t_k := \phi(k/N),
  \qquad
  k = 0, \ldots, N,
\end{equation}
where $\phi: [0,1]\to [0,T]$ is a continuously differentiable function,  which  satisfies
\begin{equation}
\label{phi0T}
    \phi(0) = 0,
    \qquad
    \phi(1) = T
\end{equation}
and has an everywhere
positive derivative $\phi'>0$.
We will refer to $\phi$ as a \emph{time discretisation  profile}. Its inverse $\phi^{-1}: [0,T]\to [0,1]$ is also a continuously differentiable function with a positive derivative
\begin{equation}
\label{psi}
    \psi(t)
    :=
    (\phi^{-1}(t))^{^\centerdot}
    =
        \frac{1}{\phi'(\phi^{-1}(t))}
\end{equation}
and satisfies
\begin{equation}
\label{invphi01}
    \phi^{-1}(0) = 0,
    \qquad
    \phi^{-1}(T) = 1.
\end{equation}
For any time $t \in [0,T]$,  the length of any of the (at most two) intervals $\Delta_k$ in (\ref{Deltak}), associated with (\ref{tttphi}) and  containing $t$,   behaves asymptotically as $\Delta t_k \sim \frac{1}{N}\phi'(\phi^{-1}(t))$, as $N \to +\infty$. Therefore, the appropriately normalised reciprocal quantity (\ref{psi})
describes the asymptotic density of the grid points (\ref{tttphi}) in the vicinity of $t$. More precisely,
\begin{equation}
\label{den}
    \lim_{N\to +\infty}
    \frac{    \#
    \{
        k = 0, \ldots, N:\
        t_k \in B
    \}
}{N}
     =
    \int_B
    \psi(t)\rd t
\end{equation}
for any Jordan measurable set $B \subset [0,T]$, where $\#(\cdot)$ is the cardinality of a set. In particular, for any interval $B:= [s,t]$, with $0 \< s \< t \< T$, the right-hand side of (\ref{den}) yields $\phi^{-1}(t)-\phi^{-1}(s) \in [0,1]$ in view of (\ref{invphi01}).
The asymptotic behaviour of the conditional covariance matrices $\Sigma_k$ in (\ref{Sigmak}), which quantify the mean-square error (\ref{MSigma}) of the numerical solution (\ref{muk}) for the linear SDE (\ref{SDElin}) over the grid (\ref{tttphi}),   is as follows.

\begin{thm}
\label{th:asy}
For a given horizon $T>0$ and a   time discretisation profile $\phi$ in (\ref{tttphi}), the conditional covariance matrices (\ref{Sigmak})
behave asymptotically as
\begin{equation}
\label{asy}
  \lim_{N\to +\infty}
  (
  N^2 \Sigma_{\lfloor N \tau \rfloor})
  =
  \Sigma(\tau)
\end{equation}
(where
$\lfloor \cdot\rfloor$ is the floor function),
with the convergence being uniform over $0 \< \tau < 1$.
Here, $\Sigma: [0,1]\to \mS_n^+$ is the solution of a Lyapunov ODE
\begin{equation}
\label{SigmaODE}
  \Sigma'(\tau)
   =
   \phi'(\tau)
   (A\Sigma(\tau) + \Sigma(\tau) A^\rT) + \phi'(\tau)^3 \mho,
\end{equation}
with the initial condition $\Sigma(0) = 0$ and the matrix $\mho \in \mS_n^+$ associated by (\ref{mho})
with the dynamics matrix $A$ from (\ref{SDElin}) and the diffusion matrix $D$ in (\ref{D}).
\end{thm}
\begin{pf}
By the variation of constants, the solution of (\ref{Sigmak}) with the initial condition in (\ref{muSigma0}),  takes the form
\begin{align}
\nonumber
    \Sigma_k
    = &
    \re^{A\sum_{\ell =0}^k\Delta t_\ell}
    \Sigma_{-1}
    \re^{A^\rT\sum_{\ell =0}^k\Delta t_\ell}\\
\nonumber
    & +
    \sum_{j=0}^k
    \re^{A\sum_{\ell ={j+1}}^k\Delta t_\ell}
    K_{\Delta t_j}
    \re^{A^\rT\sum_{\ell ={j+1}}^k\Delta t_\ell}
    (\Delta t_j)^3\\
\label{sol}
    = &
    \sum_{j=0}^k
    \re^{A(t_{k+1}-t_{j+1})}
    K_{\Delta t_j}
    \re^{A^\rT(t_{k+1}-t_{j+1})}
        (\Delta t_j)^3
\end{align}
for all $0 \< k< N$
regardless of a particular structure of the discretisation (\ref{ttt}).
Now, the grid of points (\ref{tttphi}) with a given time discretisation profile $\phi$   satisfies
\begin{align}
\label{philim}
    t_{\lfloor N\tau \rfloor}
    & = \phi(\lfloor N\tau \rfloor/N)\to \phi(\tau),\\
\nonumber
    N\Delta t_{\lfloor N\tau \rfloor}
    & =
    N
    \Big(
        \phi
        \Big(
            \frac{\lfloor N\tau \rfloor + 1}{N}
        \Big)
        -
        \phi
        \Big(
            \frac{\lfloor N\tau \rfloor}{N}
        \Big)
    \Big)\\
\label{phi'lim}
    & =
        \phi'
        \Big(
            \frac{\lfloor N\tau \rfloor + \theta}{N}
        \Big)
    \to \phi'(\tau),
    \qquad
    {\rm as}\
    N \to +\infty
\end{align}
(with $0 \< \theta \<1$ depending on $\phi$,  $N$, $\tau$) for any $\tau \in [0,1)$
because
$
    \tau - \frac{1}{N}<
    \frac{\lfloor N\tau\rfloor }{N} \< \tau
$ and the function $\phi$ is continuously differentiable.
Moreover, the convergence in (\ref{philim}), (\ref{phi'lim}) is uniform over $\tau$ since both $\phi$ and $\phi'$ are uniformly continuous on the compact interval $[0,1]$. In combination with the uniform continuity of the matrix exponential $[0,T]\ni t\mapsto \re^{tA}$  together with the boundedness of $\phi'$, the relations (\ref{sol})--(\ref{phi'lim}) and (\ref{mho}) imply that the quantity $N^2 \Sigma_{\lfloor N\tau \rfloor}$ converges uniformly over $\tau \in [0,1)$ to the same limit as the Riemann sums of the following integral:
\begin{align}
\nonumber
    \lim_{N\to +\infty}
    (N^2 \Sigma_{\lfloor N \tau \rfloor})
    & =
    \int_0^\tau
    \re^{(\phi(\tau)-\phi(v)) A}
    \mho
    \re^{(\phi(\tau)-\phi(v)) A^\rT}
    \phi'(v)^3
    \rd v\\
\label{Sigma}
    & =: \Sigma(\tau).
\end{align}
It now remains to note that the so defined function $\Sigma: [0,1]\to \mS_n^+$ satisfies the ODE (\ref{SigmaODE}) with the zero initial condition, whereby (\ref{Sigma}) establishes (\ref{asy}).
\end{pf}

As can be seen from the proof, the uniform convergence (\ref{asy}) remains valid if the floor function is replaced with the ceiling function $\lceil \cdot \rceil$ or  the function $[\,\cdot\,]$ of rounding a real number to any of the (at most two)  nearest integers. Here, we ignore a purely formal ``boundary effect'' that the quantities $\lceil  N \tau\rceil$ or $[N\tau]$ are equal to $N$ for some values of $N$ and  $\tau \in [0,1)$. This is justified by the fact that for any $\tau <1$, they are smaller than $N$ for all sufficiently large $N$.

\section{Asymptotic behaviour of mean-square error functionals}
\label{MSfun}

Theorem~\ref{th:asy} is applicable to the asymptotic behaviour of the terminal and integral mean-square error functionals (\ref{MSfin}), (\ref{MSint}).
To this end, we will use the finite-horizon observability Gramian of the pair $(A,\sqrt{M})$:
\begin{equation}
\label{Q}
  Q_t
   :=
  \int_0^t \re^{sA^\rT} M \re^{sA} \rd s,
  \qquad
  t\> 0,
\end{equation}
which satisfies the initial value problem for a Lyapunov ODE
\begin{equation}
\label{Qdot}
    \dot{Q}_t
    =
    A^\rT Q_t + Q_t A + M,
    \qquad
    Q_0 = 0,
\end{equation}
cf. (\ref{G}), (\ref{Gdot}).
Its time derivative is found from (\ref{Q}) in closed form as
\begin{equation}
\label{R}
  R_t
  :=
  \dot{Q}_t
      =
    \re^{tA^\rT} M \re^{tA}
\end{equation}
and satisfies another Lyapunov ODE
$    \dot{R}_t
    =
    A^\rT R_t + R_t A,
$ with $R_0 = M$.

\begin{thm}
\label{th:termintasy}
For a given horizon $T>0$ and a   time discretisation profile $\phi$ in (\ref{tttphi}), the mean-square error functionals (\ref{MSfin}), (\ref{MSint}) behave asymptotically as
\begin{align}
\label{asyfin}
  \lim_{N\to +\infty}
  (N^2 \cT_N)
  & =
  \int_0^T
  \frac{F_t}{\psi(t)^2}
  \rd t
  =: \Phi_T(\psi),\\
\label{asyint}
  \lim_{N\to +\infty}
  (N^2 \cI_N)
  & =
  \int_0^T
  \frac{S_t}{\psi(t)^2}
  \rd t =: \Ups_T(\psi) .
\end{align}
Here, $\psi$ is the asymptotic density (\ref{psi}) of the grid points, and
\begin{equation}
\label{FS}
  F_t := \bra \mho, R_{T-t}\ket ,
  \quad
  S_t := \bra \mho, Q_{T-t}\ket,
  \qquad
  0 \< t\< T,
\end{equation}
are nonnegative functions associated with the matrix (\ref{mho}) and the observability Gramian and its derivative from (\ref{Q})--(\ref{R}).
\end{thm}
\begin{pf}
Since $N-1 = \lfloor N \tau\rfloor$ with $\tau = 1-\frac{1}{N}$, then by the uniform convergence (\ref{asy}) and the continuity of the limit function $\Sigma$ on the interval $[0,1]$, the last conditional covariance matrix in (\ref{Sigmak}) satisfies
\begin{align}
\nonumber
    \lim_{N\to +\infty}
    (N^2 \Sigma_{N-1})
    & =
    \Sigma(1)
    =
    \int_0^1
    \re^{(T-\phi(\tau))A}
    \mho
    \re^{(T-\phi(\tau))A^\rT}
    \phi'(\tau)^3
    \rd \tau    \\
\label{Sigma1}
    & =
    \int_0^T
    \re^{(T-t)A}
    \mho
    \re^{(T-t)A^\rT}
    \frac{\rd t}{\psi(t)^2}.
\end{align}
Here, use is also made of (\ref{Sigma}) along with (\ref{phi0T}) and the integration variable change $\tau = \phi^{-1}(t)$ leading to
\begin{equation}
\label{change}
    \phi'(\tau)^3\rd \tau = \phi'(\tau)^2 \rd \phi(\tau) = \frac{\rd t}{\psi(t)^2}
\end{equation}
in view of (\ref{psi}). From (\ref{Sigma1}),
the terminal mean-square error functional (\ref{MSfin}) satisfies
\begin{align*}
    \lim_{N\to +\infty}&
    (N^2 \cT_N)
    =
    \bra M, \Sigma(1)\ket
    =
    \int_0^T
    \bra
    M,
    \re^{(T-t)A}
    \mho
    \re^{(T-t)A^\rT}
    \ket
    \frac{\rd t}{\psi(t)^2}\\
    = &
    \int_0^T
    \bra
    \re^{(T-t)A^\rT}M\re^{(T-t)A},
    \mho
    \ket
    \frac{\rd t}{\psi(t)^2}
    =
    \int_0^T
    \bra
    R_{T-t},
    \mho
    \ket
    \frac{\rd t}{\psi(t)^2},
\end{align*}
which employs the identity $\bra \alpha, \lambda \beta \rho\ket =  \bra \lambda^\rT \alpha\rho^\rT, \beta \ket$ for compatibly dimensioned real matrices $\alpha$, $\beta$, $\lambda$, $\rho$ in combination with (\ref{R}) and, by the first equality  in (\ref{FS}),  establishes (\ref{asyfin}). In a similar fashion, the uniform convergence in (\ref{phi'lim}),  (\ref{Sigma}) and the relation (\ref{change}) lead to
\begin{align}
\nonumber
    \lim_{N\to +\infty}
    &
    \Big(
    N^2
    \sum_{k=0}^{N-1}
    \Sigma_k
    \Delta t_k
    \Big)
     =
    \lim_{N\to +\infty}
    \Big(
    N^3
    \int_0^1
    \Sigma_{\lfloor N \tau\rfloor}
    \Delta t_{\lfloor N \tau\rfloor}
    \rd \tau
    \Big)\\
\nonumber
    = &
    \int_0^1
    \Sigma(\tau)
    \phi'(\tau)
    \rd \tau\\
\nonumber
    = &
    \int_0^1
    \phi'(\tau)
    \Big(
    \int_0^\tau
    \re^{(\phi(\tau)-\phi(v))A}
    \mho
    \re^{(\phi(\tau)-\phi(v))A^\rT}
    \phi'(v)^3
    \rd v
    \Big)
    \rd \tau    \\
\nonumber
    = &
    \int_0^T
    \Big(
    \int_0^t
    \re^{(t-s)A}
    \mho
    \re^{(t-s)A^\rT}
    \frac{\rd s}{\psi(s)^2}
    \Big)
    \rd t    \\
\label{Sigma2}
    =&
    \int_0^T
    \Big(
    \int_0^{T-s}
    \re^{vA}
    \mho
    \re^{vA^\rT}
    \rd v
    \Big)
    \frac{\rd s}{\psi(s)^2}.
\end{align}
In view of (\ref{mho}) and (\ref{D}),
the last inner integral on the right-hand side of (\ref{Sigma2}) is the finite-horizon controllability Gramian of the pair $(A,\frac{1}{2\sqrt{3}}AB)$ over a time interval of length $T-s$. From (\ref{Sigma2}), it follows that
the integral mean-square error functional (\ref{MSint}) satisfies
\begin{align}
\nonumber
    \lim_{N\to +\infty}
    (N^2 \cI_N)
    & =
    \int_0^1
    \bra M, \Sigma(\tau)\ket
    \phi'(\tau)
    \rd \tau \\
\nonumber
    & =
    \int_0^T
    \Big(
    \int_0^{T-s}
    \bra
    M,
    \re^{vA}
    \mho
    \re^{vA^\rT}
    \ket
    \rd v
    \Big)
    \frac{\rd s}{\psi(s)^2}\\
\nonumber
    & =
    \int_0^T
    \Big(
    \int_0^{T-s}
    \bra
    \re^{vA^\rT}M
    \re^{vA},
    \mho
    \ket
    \rd v
    \Big)
    \frac{\rd s}{\psi(s)^2}\\
\label{limint}
    & =
    \int_0^T
    \bra
    Q_{T-s},
    \mho
    \ket
    \frac{\rd s}{\psi(s)^2},
\end{align}
where use is also made of (\ref{Q}).
By the second equality in (\ref{FS}), the relation (\ref{limint}) is identical to (\ref{asyint}).
\end{pf}

Note that, in view of (\ref{Q}), (\ref{R}),  the functions (\ref{FS}) are related by
\begin{equation}
\label{FintS}
    \int_t^T F_v \rd v = S_t,
    \qquad
    0 \< t \< T.
\end{equation}
Also, by convexity of the function $0< u\mapsto \frac{1}{u^2}$ and Jensen's inequality,
$
    \frac{1}{T}
        \int_0^T
    \frac{1}{\psi(t)^2}
    \rd t
    \>
    (
    \frac{1}{T}
    \int_0^T \psi(t) \rd t)^{-2} = T^2
$,
where, in view of (\ref{psi}), (\ref{invphi01}),
\begin{equation}
\label{intpsi}
    \int_0^T \psi(t) \rd t
    =
    \phi^{-1}(T)-\phi^{-1}(0) = 1.
\end{equation}
Hence,
the right-hand side of (\ref{asyfin}) admits a lower bound
\begin{equation}
\label{T3}
    \Phi_T(\psi)
    \>
    T^3
    \min_{0\< t \< T}
    F_t
\end{equation}
for any time discretisation profile $\phi$ in (\ref{tttphi}) and thus cannot be made arbitrarily small if $F_t>0$ for all $t \in [0,T]$.

\section{$1/3$ power law in asymptotically optimal discretisation}
\label{opt}

The fine-grid asymptotic behaviour (\ref{asyfin}), (\ref{asyint}) of the mean-square error functionals (\ref{MSfin}), (\ref{MSint}), which admits an equivalent form
$
    \cT_N
    \sim
    \frac{\Phi_T(\psi)}{N^2}$ and
$    \cI_N
    \sim
    \frac{\Ups_T(\psi)}{N^2}
$,
as $N \to +\infty$,
suggests the minimisation of either of the numerators of these fractions as an optimality criterion for choosing the time discretisation profile $\phi$ in (\ref{tttphi}). This leads to similar yet different  optimal control problems
\begin{equation}
\label{probs}
    \Phi_T(\psi)
  \to \inf,
  \qquad
  \Ups_T(\psi)
  \to \inf,
\end{equation}
where the grid density $\psi: [0,T] \to (0,+\infty)$ from (\ref{psi})  plays the role of a control strategy subject to the constraint (\ref{intpsi}). The following theorem provides the solutions of the problems  (\ref{probs}).

\begin{thm}
\label{th:opt}
Suppose the dynamics, diffusion and weighting matrices $A$, $D$, $M$ in (\ref{SDElin}), (\ref{D}),  (\ref{MSigma}) satisfy
\begin{equation}
\label{good}
    \det
    (\bra M, A^j D (A^\rT)^k\ket)_{1\< j,k \< n} >0.
\end{equation}
Then the functional $\Phi_T$ in (\ref{asyfin}) achieves its minimum value
\begin{equation}
\label{minfin}
  \min_\phi
  \Phi_T(\psi)
  =
  \Big(
  \int_0^T
  \sqrt[3]{F_t}
  \rd t
  \Big)^3
\end{equation}
at a unique time discretisation profile $\phi$  with the inverse
\begin{equation}
\label{phifin}
    \phi^{-1}(t)
    =
  \frac{\int_0^t\sqrt[3]{F_u} \rd u}{\int_0^T \sqrt[3]{F_v} \rd v},
  \qquad
  0 \< t \< T,
\end{equation}
where (\ref{FS}) is used.
Similarly, the functional $\Ups_T$ in (\ref{asyint}) achieves its infimum
\begin{equation}
\label{minint}
  \inf_{\phi}
  \int_0^T
  \frac{S_t}{\psi(t)^2}
  \rd t
  =
  \Big(
  \int_0^T
  \sqrt[3]{S_t}
  \rd t
  \Big)^3
\end{equation}
in an extended class of functions $\phi$, which are continuous on $[0,T]$ and  continuously differentiable  on $[0,T)$, with $\phi'>0$ on $[0,T)$,   and satisfy the boundary conditions (\ref{phi0T}),  with the inverse
\begin{equation}
\label{phiint}
    \phi^{-1}(t)
    =
  \frac{\int_0^t\sqrt[3]{S_u} \rd u}{\int_0^T \sqrt[3]{S_v} \rd v},
  \qquad
  0 \< t \< T.
\end{equation}
\end{thm}
\begin{pf}
In view of (\ref{D}), the condition (\ref{good}) means nonsingularity of the Gram matrix for the matrices
$
    \sqrt{M}A^k B \in \mR^{n\x m}$, with
    $k = 1, \ldots, n$,  and hence, is equivalent to their linear independence. By (\ref{mho}) and   the Cayley-Hamilton theorem,  this implies that (\ref{R}) satisfies
\begin{equation}
\label{mhoR}
    \bra \mho,  R_t \ket  = \bra \mho, \re^{tA^\rT} M \re^{tA}\ket = \frac{1}{12}\|\sqrt{M} \re^{tA}AB \|^2 > 0
\end{equation}
(with $\|\cdot\|$ the Frobenius norm of matrices) for any $t \in \mR$,  and thus, $F_t>0$ in (\ref{FS})   for all $0 \< t \< T$. In turn, the latter, combined with (\ref{FintS}), implies that $S_t >0$ for all $0 \< t < T$.
Now, for any time discretisation profile $\phi$ in (\ref{tttphi}),  application of H\"{o}lder's inequality with $p:= 3$ and $q:= \frac{3}{2}$ (so that $\frac{1}{p} + \frac{1}{q} = 1$) leads to
\begin{align}
\nonumber
    \int_0^T
    \sqrt[3]{F_t}
    \rd t
    & =
    \int_0^T
    \sqrt[3]{\frac{F_t}{\psi(t)^2}}\,
    \psi(t)^{2/3}
    \rd t\\
\label{Fint}
    & \<
    \sqrt[3]{\Phi_T(\psi)}
    \Big(
    \int_0^T
    \psi(t)
    \rd t
    \Big)^{2/3},
\end{align}
where  $\Phi_T(\psi)$ is the functional (\ref{asyfin}) of the grid density   $\psi$ from (\ref{psi}). Since the latter  satisfies (\ref{intpsi}),
then (\ref{Fint}) implies that
$
    \Phi_T(\psi)
    \>
    (
        \int_0^T
    \sqrt[3]{F_t}
    \rd t
    )^3
$,
with the inequality holding as an equality for a unique function $\psi> 0$  subject to the constraint (\ref{intpsi}):
\begin{equation}
\label{psiopt}
  \psi(t)
  =
  \frac{\sqrt[3]{F_t}}{\int_0^T
    \sqrt[3]{F_s}
    \rd s},
    \qquad
    0 \< t \< T.
\end{equation}
This establishes (\ref{minfin}), with the optimal time discretisation profile $\phi$ being recovered from (\ref{psiopt}) according to (\ref{psi}), (\ref{phifin}). The proof of (\ref{minint}) is similar to the above and yields the optimal grid density
\begin{equation}
\label{psiopt2}
  \psi(t)
  =
  \frac{\sqrt[3]{S_t}}{\int_0^T
    \sqrt[3]{S_v}
    \rd v},
    \qquad
    0 \< t \< T,
\end{equation}
leading to the time discretisation profile (\ref{phiint}). However,  while $\psi$ in (\ref{psiopt2}) is positive on $[0,T)$, it vanishes at the terminal time:
\begin{equation}
\label{psiT0}
    \psi(T)= 0
\end{equation}
because $S_T = \bra \mho, Q_0\ket = 0$ in view of (\ref{FS}) and the zero initial condition in (\ref{Qdot}).
\end{pf}

Note that the right-hand side of (\ref{T3}), as a lower bound for the exact minimum value (\ref{minfin}), is conservative if $F_t$ varies substantially over $t \in [0,T]$.  Also, the fact that the asymptotically optimal density $\psi$ for the integral mean-square error  functional (\ref{MSint}) satisfies (\ref{psiT0}) manifests the vanishing value of near-terminal grid points for the integral performance of the numerical solution (\ref{muk}) over the interval $[0,T]$. From (\ref{Qdot}), (\ref{FS}), it follows that $Q_t\sim tM$,  as $t\to 0+$, and hence,   $S_t \sim \bra \mho, M\ket(T-t)$,  as $t\to T-$, where $\bra \mho, M\ket = \frac{1}{12} \|\sqrt{M}AB\|^2>0$  in view of  (\ref{D}), (\ref{mho}), (\ref{good}). Therefore, the grid density (\ref{psiopt2}) has the following near-terminal asymptotic  behaviour:
\begin{equation}
\label{root3}
    \psi(t)
    \sim
    C
    \sqrt[3]{T-t},
    \quad
    {\rm as}\
    t \to T-,
    \quad
    C:=     \frac{\sqrt[3]{\frac{1}{12} \|\sqrt{M}AB\|^2}}{\int_0^T
    \sqrt[3]{S_v}
    \rd v},
\end{equation}
where $C$ is a constant coefficient.
The relation (\ref{psiT0})   and its refinement
(\ref{root3}) can also be understood from the viewpoint of dynamic programming which provides another (though more complicated in this case) way to arrive at the solutions of the constrained optimization problems (\ref{probs}) in Theorem~\ref{th:opt}. In fact, the latter can be ``reverse engineered'' for a hint on the structure of the corresponding Bellman functions. Also note that, as can be seen from the proofs of Theorems~\ref{th:asy}--\ref{th:opt},  the ``$1/3$'' power laws in (\ref{minfin})--(\ref{phiint}) and (\ref{root3}) originate from the $(\Delta t_k)^3$ terms in (\ref{Sigmak}).

\section{One-dimensional illustrative example}
\label{example}
\vspace{-2mm}
As an illustration,
consider an Ornstein-Uhlenbeck (OU) process $X$ governed by (\ref{SDElin}) with $n = m = 1$, $A< 0$ and $B\in \mR\setminus \{0\}$, so that $D= B^2 >0$ in (\ref{D}) and $\mho = \frac{1}{12} A^2 D>0$ in (\ref{mho}).  Also, we let $M=1$ in (\ref{MSigma})--(\ref{MSint}) without loss of generality.  Then (\ref{R}), (\ref{FS}), (\ref{mhoR}) lead to the following   minimum value in (\ref{minfin}):
\begin{align}
\nonumber
    \min_\phi \Phi_T(\psi)
    & =
    \frac{1}{12}
    A^2D
    \Big(
    \int_0^T
    \sqrt[3]{
    \re^{2A(T-t)}}
    \rd t
    \Big)^3\\\
\label{minPhi}
    & =
    \frac{9D}{32 A}
    (\re^{\frac{2}{3}AT}-1)^3
    =
    \frac{9}{16}
    G_\infty
    (1-\re^{\frac{2}{3}AT})^3    ,
\end{align}
where, in accordance with (\ref{G}), (\ref{Gdot}),   $G_\infty := \lim_{t\to +\infty} G_t = -\frac{D}{2A} = \lim_{t\to +\infty} \bE (X_t^2)$ is the infinite-horizon controllability Gramian of $(A,B)$ in the scalar case being considered and coincides with the invariant variance of the OU process. The asymptotically optimal grid density (\ref{psiopt}), which delivers (\ref{minPhi}),  is given by
\begin{equation}
\label{psiopt3}
    \psi(t)
    =
    \frac{\re^{\frac{2}{3} A (T-t)}}{\int_0^T \re^{\frac{2}{3} A s}\rd s}
    =
    \frac{2A}{3}
    \frac{\re^{\frac{2}{3}A(T-t)}}{\re^{\frac{2}{3}AT}-1},
    \qquad
    0\< t \< T,
\end{equation}
and grows exponentially fast over the time interval $[0,T]$, with
$
    \frac{\psi(T)}{\psi(0)} = \re^{-\frac{2}{3} AT}
$.
Compared to the uniform grid with the constant density $\frac{1}{T}$ and (\ref{asyfin}) leading to
$
    \Phi_T(1/T)
    =
    T^2
    \int_0^T
    F_t \rd t
    =
    \frac{1}{12}
    (AT)^2
    D
    \int_0^T
    \re^{2A(T-t)}
    \rd t
    =
    \frac{1}{12}
    G_\infty(AT)^2(1-\re^{2AT})
$,  the advantage of its optimal counterpart (\ref{psiopt3}) is described by the ratio
$
    \frac{\Phi_T(1/T)}{\min_\phi \Phi_T(\psi)}
    =
    \frac{4}{27}
    (AT)^2
    \frac{1-\re^{2AT}}{(1-\re^{\frac{2}{3} AT})^3}
$,
which is large when $T \gg -\frac{1}{A}$ (that is,  at horizons significantly exceeding the correlation time of the OU process) and behaves asymptotically as
    $\frac{4}{27}
    (AT)^2$, as $T \to +\infty$.

\vspace{-2mm}
\section{Conclusion}
\label{conc}
\vspace{-3mm}

We have revisited the numerical integration of SDEs as a filtering problem where the strong solution  is recursively estimated based on the increments of the driving Wiener process as discrete-time  observations. We have applied Kalman filtering to a mean-square optimal numerical solution of linear SDEs with arbitrary time grids.  Its fine-grid asymptotic accuracy has been discussed for a class of monotonic transformations of uniform grids. We have obtained asymptotically optimal discretisation profiles to minimise the terminal and integral mean-square error functionals and found $1/3$ power laws in their structure. The stochastic filtering and asymptotic grid optimisation approach to  numerical integration can, in principle,   be extended to time-varying and nonlinear SDEs.

\end{document}